\documentclass[prc,aps,showpacs,twocolumn]{revtex4}
\usepackage{amsmath}
\usepackage{amssymb}
\usepackage{graphicx,color}

\begin{document}

\title{Benchmark calculation of inclusive electromagnetic responses\\ in the four-body nuclear system}

\author{Ionel Stetcu$^{1}$}
\altaffiliation{On leave from the National Institute for Physics 
and Nuclear Engineering ``Horia Hulubei", Bucharest, Romania.
{\tt email: istet@email.arizona.edu}}
\author{Sofia Quaglioni$^{1}$}
\altaffiliation{\tt email: sofia@physics.arizona.edu}
\author{Sonia Bacca$^{2}$}
\altaffiliation{\tt email: s.bacca@gsi.de}
\author{Bruce R. Barrett$^1$} 
\author{Calvin W. Johnson$^{3}$} 
\author{Petr Navr\'atil$^{4}$} 
\author{Nir Barnea$^{5}$}
\author{Winfried Leidemann$^{6}$} 
\author{Giuseppina Orlandini$^{6}$}

 \affiliation{$^{1}$Department of Physics, University of Arizona, P.O. Box 210081, Tucson,
 Arizona 85721,\\
 $^{2}$Gesellschaft f\"{u}r Schwerionenforschung, Planckstr.~1,
 64291 Darmstadt, Germany,\\
 $^{3}$Physics Department,
 San Diego State University,
 5500 Campanile Drive, San Diego, California 92182-1233,\\
 $^{4}$Lawrence Livermore National Laboratory, Livermore, P.O. Box 808,
 California 94551,\\
 $^{5}$Racah Institute of Physics, The Hebrew University, 91904, Jerusalem,
 Israel,\\
 $^{6}$ Dipartimento di Fisica, Universit\`{a} di Trento and INFN\\
 (Gruppo Collegato di Trento), via Sommarive 14, I-38050 Povo, Italy}
 
\date{\today{}}

\begin{abstract}
Both the no-core shell model and the effective interaction hyperspherical harmonic approaches are applied to the calculation of different response
functions to external electromagnetic probes, using the Lorentz integral transform method. 
The test is performed on the four-body nuclear system, within a simple potential model.
The quality of the agreement in the various cases is discussed, together with the perspectives for rigorous \emph{ab initio} calculations of cross sections of heavier nuclei. 
\end{abstract}
\maketitle
\bigskip

\section{introduction}
A challenging problem in nuclear physics is to calculate nuclear properties microscopically, 
using realistic nuclear forces. For light systems 
($A=2,3$), the Schr\"odinger equation can be solved with a high degree of accuracy 
and both ground state properties and reaction cross sections have been 
calculated~\cite{AS:1991,Carlson:1998,Arenhovel:2005,marcucci:014001,skibinski:044002,Golak:41589,deltuva:054004}.

For heavier systems, the Green's function Monte Carlo (GFMC) method 
\cite{Pieper:2002ne,pieper:054325} as well as the no-core shell
model (NCSM) approach~\cite{Navratil:2000ww} have been successfully applied
for the \emph{ab initio} description of nuclear properties, using 
realistic nucleon-nucleon (NN) and three-nucleon forces.
GFMC calculations, while more accurate, are limited to 
masses up to $A=12$, while the NCSM can handle a larger range of masses,
up to $A=16$ and beyond~\cite{A48,ne20NCSM}.

However, for A$>3$ a big obstacle is encountered if one wants to calculate reaction observables involving states in the continuum, because of the enormous difficulties in calculating many-body scattering states.

Such difficulties can be avoided using  
integral transforms, which reduce the continuum problem to a bound-state-like problem~\cite{EFROS:1985,EFROS:1993,EFROS:1999}, so that only bound state techniques are required. While not all integral transforms are appropriate for a practical
use~\cite{Carlson:1992,Efros:1993fbs} (the inversion of the transforms is an ``ill posed problem''~\cite{Tikonov:1977}), the integral transform with a Lorentz kernel~\cite{Efros:1994iq} appears to be the practical tool for such calculations. In fact, it has allowed the calculation of electromagnetic reaction cross sections beyond break-up thresholds of nuclei from 
$A=3$~\cite{Efros:011002} to A=7~\cite{Bacca:2004dr}. 
Such calculations have also been possible thanks to the use of the effective interaction hyperspherical harmonic (EIHH)~\cite{barnea:054001,barnea:054003} approach, a very accurate bound state technique, which, similar to the NCSM approach, uses the concept of an effective interaction to speed up the convergence of the basis expansions. In particular,
by means of the Lorentz integral transform (LIT) method combined with the EIHH technique, total photoabsorption cross sections of six- \cite{Bacca:2001kr,bacca:057001} and seven-body nuclei~\cite{Bacca:2004dr}
with semirealistic NN potentials have been calculated. In a very recent work~\cite{gazit:112301} the $^4$He total photoabsorption cross section could even be calculated with a realistic nuclear force (two- and three-body potential). Moreover, a similar formalism has been recently applied to describe exclusive electromagnetic processes of the four-body 
nuclear system with a semirealistic potential~\cite{quaglioni:044002,quaglioni:064002,andreasi:06eps}.

While the EIHH and NCSM approaches are rather similar, only the latter has made use of realistic interactions in calculations with $A>4$ . Indeed, NCSM has the advantage that one can use an equivalent Slater determinant basis, allowing
description of a larger range of masses. Therefore, it is of great interest
 to investigate the possibility of applying the NCSM to the LIT equations.
In fact, if one could obtain reliable results in this way, the possibility to study also reactions on heavier nuclei by means of realistic \emph{ab initio} approaches would open up.

The purpose of the present work is to investigate the applicability of the NCSM approach to the solutions of the bound-state equations required by the LIT method. 
The similarities between the EIHH and the NCSM makes the task straightforward, in principle. On the other hand, the practical implementation of the method, especially concerning the problems of convergence, might lead to difficulties. 

In the following, we present the results of a test consisting in calculating, within the EIHH and NCSM approaches, the LIT of the four-body response functions to two different  excitation operators. Because of the convergence properties, the input interaction is the simple Minnesota (MN)~\cite{Minnesota} potential. Two operators of different nature and range have been chosen, \emph{i.e.}, the isovector dipole and isoscalar quadrupole.

The paper is organized as follows.
In Sec. \ref{Sec:Overview}, the theoretical approaches are discussed, that is the LIT as well as the NCSM and EIHH methods, respectively. Results are presented in Sec. \ref{Sec:Results}, while conclusions and perspectives are discussed in Section \ref{Sec:Conclusions}.

\section{Theoretical overview}
\label{Sec:Overview}

\subsection{The LIT approach}

The inclusive cross sections of reactions induced by perturbative external probes
are generally written in terms of the so-called response functions defined as:
\begin{equation}
{\label{1}R(\omega)=\int d\Psi_{f}\left|\left\langle \Psi_{f}\right|\hat{O}\left|\Psi_{0}\right\rangle \right|^{2}\delta(E_{f}-E_{0}-\omega)}\,,
\end{equation}
where  $\omega$ represents the
energy transferred by the probe, and 
 $\hat O$ the excitation operator. Wave functions and energies of the ground
and final states of the perturbed system are denoted by $\left|\Psi_{0/f}\right>$ and $E_{0/f}$, respectively.

In the LIT method~\cite{Efros:1994iq}
one obtains $R(\omega)$ after the inversion of an integral transform
with a Lorentzian kernel 
\begin{equation}
L(\sigma_{R},\sigma_{I})=\int d\omega\frac{R(\omega)}{(\omega-\sigma_{R})^{2}+\sigma_{I}^{2}}=\langle \widetilde{\Psi}|\widetilde{\Psi}\rangle \,.\label{2}
\end{equation}
 The state $\widetilde{\Psi}$ is the unique
solution of the inhomogeneous ``Schr\"{o}dinger-like'' equation
\begin{equation}
(H-E_{0}-\sigma_{R}+i\sigma_{I})|\widetilde{\Psi}\rangle=\hat{O}|{\Psi_{0}}\rangle.\label{3}
\end{equation}
Because of the presence of an imaginary part $\sigma_I$ in Eq.~(\ref{3}) 
and the fact that the right-hand side of this same equation is localized,
one has an asymptotic
boundary condition similar to a bound state. Thus, one can apply bound-state
techniques for its solution, and, in particular, expansions over basis sets 
of localized functions.
In the present paper we solve Eq.~(\ref{3}) using both the NCSM and
EIHH methods, which will be briefly described below.

In both cases we evaluate the LIT by calculating the quantity
$L(\sigma)=\langle\widetilde{\Psi}|\widetilde{\Psi}\rangle$ 
directly~\cite{Marchisio:2003} via the
Lanczos algorithm. Hence, one finds that
the LIT can be written as a continuous fraction 
\begin{equation}
L(\sigma)=\frac{M_0}{\sigma_{I}}
~{\rm Im}\frac{1}{(z-a_{0})-\frac{b_{1}^{2}}{(z-a_{1})-\frac{b_{2}^{2}}{(z-a_{2})-b_{3}^{2}....}}}\,,\label{7}
\end{equation}
in terms of the Lanczos coefficients $a_{n}$ and $b_{n}$, where
$\sigma=\sigma_R+i\sigma_I$ and $z=E_{0}+\sigma$. The quantity 
\begin{equation}
M_{0}=\langle\Psi_{0}|\hat{O}^{\dagger}\hat{O}|\Psi_{0}\rangle
\end{equation}
is the zero-th moment of the distribution $R(\omega)$ and, therefore,
equivalent to the total strength induced by the excitation operator 
(non-energy weighted sum rule). A parameter-free smooth response function 
$R(\omega)$ is obtained by inversion of the integral transform of Eq.~(\ref{2})
(for inversion methods see Refs.~\cite{Efros:1999fbs} and~\cite{andreasi:05eps}).

Here we would like to note that similar applications have already been used in the 
past~\cite{Caurier:1990dc,Caurier:1994xg} for evaluating 
response functions by means of shell-model calculations in terms of the so-called ``Lanczos 
response'' (LR). In these cases, LR is
\begin{equation}
LR(\omega)=M_0\sum_{n=1}^{\mathcal N} S_n~\delta(E_n-E_0-\omega)\,,
\label{LRdelta}
\end{equation}
where ${\mathcal N}$ is the number of Lanczos iterations, while $E_n$ 
and $S_n$ are the eigenvalues and transition strengths obtained with
the Lanczos algorithm, respectively. Starting with Eq.~(\ref{LRdelta}), one usually replaces 
the $\delta$ function by a resolution function
(Lorentzian or Gaussian) with a width parameter appropriate to the 
experiment (a typical value is $\sigma_I=0.25$ MeV)~\cite{haxton:065501}, or one evaluates the running integral
\begin{equation}
I(\omega)=\int_{E_{th}}^\omega LR(\omega^\prime)\,d\omega^\prime\,,
\end{equation}
where $E_{th}$ is the break-up threshold energy of the system.
While the LR has the same underlying physical content as
$L(\sigma_R,\sigma_I)$, since in both approaches one
starts from the poles and the residues of the Green's function,
it is evident that the information is processed in a different way.
In fact, in the LIT method
Eq.~(\ref{7}) is regarded as an integral transform of the response function,
which  is then recovered via numerical inversion. In this approach, one
typically chooses larger values of $\sigma_I$ (~$10-20$ MeV),
of the order of the strunctures expected in the response function,
such that $L(\sigma_R, \sigma_I)$ can be calculated with the
sufficient numerical precision that allows a stable
 inversion. 
After the inversion one recovers a smooth
response, which is independent of $\sigma_I$ in a large range of
values for this parameter.

\subsection{The NCSM and the EIHH approaches}

The EIHH and NCSM approaches are spectral resolution methods, where
one performs an expansion of the Schr\"{o}dinger wave function in
terms of a complete set of basis states: the hyperspherical-harmonics
(HH) functions in the EIHH case, and the HO
basis functions in the NCSM case. Obviously, due to computational
limitations, the basis has to be truncated. The truncated set of states
forms the so-called model or $P$-space, and the excluded basis functions
build its complementary $Q$-space, such that the sum $P+Q=1$ spans
the total Hilbert space. It is well-known that for a given NN
interaction the HO and the HH expansions converge rather slowly. In
order to accelerate the convergence pattern, both methods make use
of an effective interaction introduced by Navr\'atil 
\textit{et. al.}~\cite{Navratil:1996vm,Navratil:1997qb}, which follows a unitary 
transformation approach.
For the finite $P$-space, the corresponding effective interaction is
obtained by means of a unitary
transformation~\cite{DaProvidencia:1964,Suzuki:1980,Suzuki:1982,Suzuki:1983}. 
In general, the exact effective
interaction for an $A$-body system will have irreducible $A$-body
matrix elements, even if the bare interaction is only two-body, and
its construction requires the exact solution of the initial problem.
Since the solution for the $A$-body problem is the primary goal of
the approach, an exact calculation of the effective interaction is
not practical, and one has to resort to approximations. Thus, the effective
interaction is approximated as a sum of $a$-body effective interaction
terms ($a<A)$, derived from the solution of the $a$-body problem.
This procedure is called the cluster approximation, and in practice,
we use $a=2$ or $3$. The effective Hamiltonian is then obtained
by replacing the bare interaction in the $A$-body Hamiltonian by
the effective one. However, because of the approximation introduced
in order to compute the effective interaction, the resulting eigenvalues
of the effective Hamiltonian do not exactly reproduce the values in
the full space, the signature being a dependence of the observables
on the parameters of the model space. 
For a given cluster approximation, the eigenenergies
converge to the bare ones by increasing the size of the model space.
And in the limit $P\longrightarrow1$ this procedure becomes exact
\cite{mintkevich:044005}. Alternately, one can obtain convergence in a fixed
model space by increasing the cluster size, as in this case the approximate
effective interaction converges to the exact one.

In the NCSM, one works in an HO basis, and the
$P$-space is spanned by states with the total number of HO quanta
$N\leq N_{\textrm{max}}$. For the purpose of this paper, we neglect
three-body forces and use local interactions, although the non-local
interactions do not present an impediment~\cite{Navratil:1999pw}. Therefore, the 
intrinsic Hamiltonian
describing a system of $A$ nucleons is simply
\begin{equation}
H=\frac{1}{A}\sum_{i<j=1}^{A}\frac{(\vec{p}_{i}-\vec{p}_j)^2}{2m}+\sum_{i<j=1}^{A}V(\vec{r}_{i}-\vec{r}_{j})
\label{intham}
\end{equation}
 with $\vec{p}_{i}$ and $\vec{r}_{i}$ being the momentum and the
position vectors of the \emph{i}th particle, respectively; $V$ being
the NN potential; and $m$ the nucleon mass. 
%
%
In order to facilitate the use of the convenient HO basis for evaluating the effective interaction, Eq.~(\ref{intham}) is modified by adding an HO center-of-mass Hamiltonian $\vec{P}^2/(2Am)+Am\Omega^2\vec{R}^2/2$,
\begin{equation}
\vec{P}=\sum_{i=1}^A\vec{p}_i\quad{\rm and}\quad \vec{R}=\frac{1}{A}\sum_{i=1}^A\vec{r}_i\,.
\end{equation}
Thus, the $A$-body Hamiltonian takes the form 
\begin{eqnarray}
H_{A}^{\Omega} & = & \sum_{i=1}^{A}\left[\frac{\vec{p}_{i}^{~2}}{2m}+\frac{1}{2}m\Omega^{2}\vec{r}_{i}^{~2}\right]\nonumber\\
 & + & \sum_{i<j=1}^{A}\left[V(\vec{r}_{i}-\vec{r}_{j})
-\frac{m\Omega^{2}}{2A}(\vec{r}_{i}-\vec{r}_{j})^{2}\right]\,,
\end{eqnarray}
 where $\Omega$ is the HO parameter. As the NN potential depends
on the relative coordinates, the added center-of-mass HO term has
no influence on the internal motion in the full space.

In the present NCSM calculations, we use the three-body cluster approximation for the effective interaction.
In order to derive it, one solves the three-body problem for the cluster Hamiltonian
\begin{eqnarray}
H_{3}^{\Omega} & = & \sum_{i=1}^{3}\left[\frac{\vec{p}_{i}^{~2}}{2m}+\frac{1}{2}m\Omega^{2}\vec{r}_{i}^{~2}\right]\nonumber\\
 & + & \sum_{i<j=1}^{3}\left[V(\vec{r}_{i}-\vec{r}_{j})
-\frac{m\Omega^{2}}{2A}(\vec{r}_{i}-\vec{r}_{j})^{2}\right]\,\,,
\label{HOmega3}
\end{eqnarray}
and determines the three-body unitary transformation $X^{(3)}$ from the condition
that the transformed Hamiltonian
\begin{equation}
{\mathcal H}^{(3)}=X^{(3)~-1}H_{3}^{\Omega}X^{(3)}
\end{equation}
preserves the solutions of Eq.~(\ref{HOmega3}) in a subspace $P^{(3)}$ of the full three-body Hilbert space.
This approximation introduces dependence on the oscillator parameter 
$\Omega$ so that one has to search for a range of $\Omega$-values that minimizes such a dependence.
The effective Hamiltonian for the $A$-body system in the three-body cluster approximation is then defined as
\begin{eqnarray}
{\mathcal H}_{eff} & = & \sum_{i=1}^{A}\left[\frac{\vec{p}_{i}^{~2}}{2m}+\frac{1}{2}m\Omega^{2}\vec{r}_{i}^{~2}\right]\nonumber\\
 &  & +\frac{1}{A-2}\sum_{i<j<k=1}^{A}{\mathcal V}^{(3)}_{ijk}\,\,,
\end{eqnarray}
with
\begin{equation}
{\mathcal V}^{(3)}={\mathcal H}^{(3)}-\sum_{i=1}^{3}\left[\frac{\vec{p}_{i}^{~2}}{2m}+\frac{1}{2}m\Omega^{2}\vec{r}_{i}^{~2}\right]\,\,.
\end{equation}
Note that, because of the decoupling condition
\begin{equation}
 Q^{(3)}{\mathcal H}^{(3)}P^{(3)}=0\,\,,
\end{equation}
the three-body effective interaction is energy-independent. A more detailed description of the derivation of the operator $X^{(3)}$ can be found in Refs.~\cite{Navratil:1999pw,Navratil:2000gs,Navratil:152502,Nogga:2005hp}. 

In a consistent approach, one should use effective transition operators,
derived by means of the same transformation that determines the effective
interaction. While the renormalization at the three-cluster level
has not been investigated for general operators, studies of the latter in the two-body
cluster approximation have shown little effect of the renormalization for long-range
observables \cite{Stetcu:2004wh,stetcu:037307}.  
Although it is conceivable that a renormalization at the
three-body cluster level can show some improvement for long-range operators in smaller
model spaces, this task is extremely demanding and not justified since we obtain model 
space independent results in the largest model spaces used in this work.

The present four-nucleon $P$-space calculations are performed in a properly 
antisymmetrized translationally invariant HO basis, as described in~\cite{Navratil:1999pw},
where the interested
reader can find details on how to compute an one-body translationally invariant
operator in such a basis.

In the EIHH method~\cite{barnea:054001,barnea:054003} the calculation is performed with the HH basis
and the spaces $P$ and $Q$ are defined
by means of the hyperspherical quantum number $K$. Thus, the model space
includes all the HH functions with $K\leq K_{\mathrm{max}}$. The
intrinsic Hamiltonian in the hyperspherical coordinates is given by 
\begin{equation}
H=\frac{1}{2m}\left[-\Delta_{\rho}+\frac{\hat{K}^{2}}{\rho^{2}}\right]+
\sum_{i<j=1}^{A}V(\vec{r}_{i}-\vec{r}_{j})\;,\label{HH1}
\end{equation}
where $\rho$ is the hyperradius and $\Delta_{\rho}$ contains derivatives
with respect to $\rho$ only. The grand-angular momentum operator
$\hat{K}^{2}$ is a function of the variables of particles $A$ and
$(A-1)$ and of $\hat{K}_{A-2}$, the grand angular momentum operator
of the residual subsystem \cite{Efros:1972}. Unlike in the case of NCSM
calculation, in this work we use a two-body cluster approximation for
EIHH, which already yields an excellent convergence pattern.
The starting point for the construction of the effective interaction is a two-body-like Hamiltonian, 
\begin{equation}
H_{2}(\rho)=\frac{1}{2m}\frac{\hat{K}^{2}}{\rho^{2}}+V(\vec{r}_{A,A-1})\;,
\label{HH2}
\end{equation}
 where $\vec{r}_{A,A-1}$ is the relative coordinate between particles
$A$ and $A-1$, but the total hyperspherical kinetic energy is considered.
Due to the presence of the hyperradius $\rho$, which is a collective
coordinate, the resulting effective interaction produces a sort of ``medium correction''.
Moreover, since it depends also on the $K_{A-2}$ quantum number of the $A-2$ particle 
subsystem, it is state dependent. The sum of these leads to a faster rate of convergence.
 The HH functions are constructed starting from  
the relative Jacobi coordinates, thus the  center of mass motion is
removed from the very beginning. 
The Fermionic basis states are obtained by coupling spin-isospin states, \emph{i.e.}, states  
that belong to irreducible representations
of the permutation group $S_{A}$, with spatial HH states to yield completely antisymmetric 
basis functions. The basis is constructed by using
the powerful algorithms of 
\cite{Barnea:1997,Barnea:1998,Novoselsky:1994,Novoselsky:1995,Barnea:1999}.

\section{Results}
\label{Sec:Results}

The aim of the present work is to investigate the reliability of the NCSM 
approach for the description of inclusive response functions via the LIT 
method, and not to give a realistic prediction of those response functions. 
Therefore, although applications of the NCSM and the EIHH have already been 
performed and published using realistic NN forces as well as theoretical 
three nucleon forces, for the purpose of our comparison it is more convenient 
to employ the semi-realistic MN~\cite{Minnesota} potential. 
The advantage of the MN potential is that it consists of Gaussian-type potentials 
and has a rather soft core, leading to good convergence rate for both the NCSM and the EIHH.

We present results for two operators different in isospin nature and range, 
giving the leading contribution to low-energy electromagnetic reactions. These are the isovector 
dipole and the isoscalar quadrupole operators, respectively:
\begin{eqnarray}
\hat D&=&\sqrt{\frac{4\pi}{3}}\sum_{i=1}^A \frac{\tau^z_i}{2} r_i  Y_{10}(\hat{r}_i)\,,\\
\hat Q&=&\sqrt{\frac{16\pi}{5}}\sum_{i=1}^A \frac{1}{2}r^2_i Y_{20}(\hat{r}_i)\,.
\end{eqnarray}

The NCSM calculations are also carried out for four different 
HO frequencies ($\hbar\Omega=12, 19, 28$ and $40$ MeV), in order to study the 
dependence of the resulting LIT on this parameter and to provide an estimate
for the theoretical uncertainties.

\begin{figure}[t]
\includegraphics*[scale=0.65]{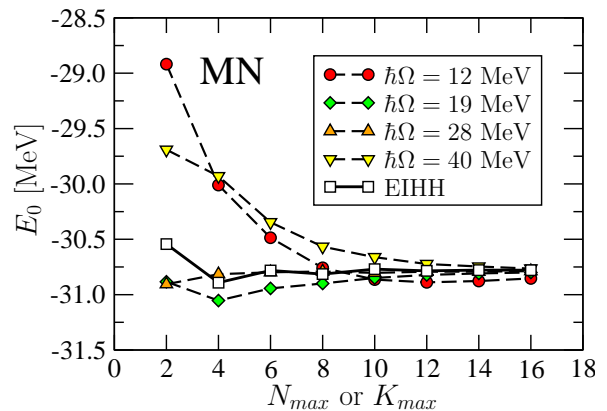}
\caption{(Color online) $^4$He ground-state energy for the MN potential 
as a function of the HO excitations allowed in the NCSM model space $N_{max}$ and 
of the maximal value of the HH grand-angular momentum quantum 
number $K_{max}$ in the EIHH expansion.} 
\label{figure1}
\end{figure}

While not the goal of the present calculation, we illustrate the convergence 
patterns for the $^4$He ground-state energy in Fig.~\ref{figure1}. This convergence 
is important because (i) the ground-state wave function $|\Psi_0\rangle$  enters 
the source term in the LIT equation (\ref{2}), and (ii) the total strength $M_0$
depends entirely on $|\Psi_0\rangle$.
A good convergence is reached already at $K_{max}=10$ for EIHH, whereas one 
obtains slower or faster convergence for NCSM, depending on the value of the 
HO frequency $\Omega$. In particular, one finds a weak dependence on the model 
space size for $\hbar\Omega=19$ and $28$ MeV. The five best results, which occur for
$N_{max}=16$ and $K_{max}=16$, 
all agree within $0.2\%$. Note that the Coulomb interaction is not included, 
so that the $^4$He g.s. energy obtained in our calculations  differs from previously
published results for the MN potential ($-29.96(1)$ MeV~\cite{Barnea:1999be}).

\begin{table}[t]
\caption{Summary of $^4$He ground-states properties obtained using the MN~\cite{Minnesota} 
potential model without Coulomb interaction. 
The evaluation of total strength of the isoscalar quadrupole 
transition, $M_0(\hat Q)$, for the NCSM does not take into account the $\hbar\Omega=40$ MeV result, 
which is not yet converged.}
\label{summary}
  \begin{ruledtabular}
\begin{tabular}{c c c c}
&$E_0$ [MeV]&$M_0(\hat D)$ [fm$^2$]& $M_0(\hat Q)$ [fm$^4$]\\
\hline
EIHH &$-30.779(1)$ & $0.7883(1)$ & $8.54(1)\phantom{4}$\\
NCSM &$-30.80(5)\phantom{1}$  & $0.786(6)\phantom{1}$  & $8.56(14)$\\
\end{tabular}
  \end{ruledtabular}
\end{table}

In Table \ref{summary}, we compare the $^4$He ground state properties obtained using the MN
two-body interaction in both EIHH and NCSM. One finds an excellent agreement for the binding
energy, as well as dipole and quadrupole total strengths, which will be discussed in more detail
below.

\subsection{Dipole response}

\begin{figure}[b]
\includegraphics*[scale=0.65]{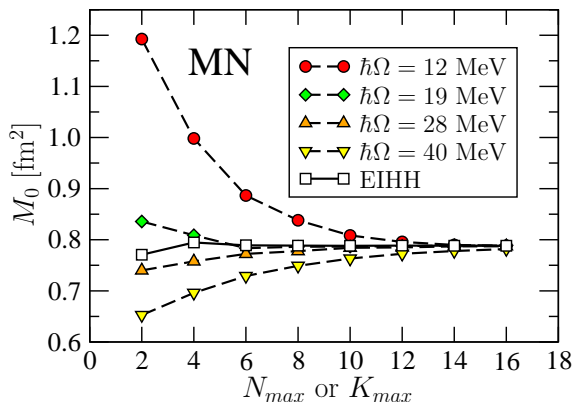}
\caption{(Color online) $^4$He total dipole strength for the MN potential as a 
function of the HO excitations allowed in the NCSM model space $N_{max}$ and of the maximal 
value of the HH grand-angular momentum quantum number $K_{max}$ in the EIHH expansion.} 
\label{figure2}
\end{figure}

We start the discussion with the results for the isovector dipole transition.
In Fig.~\ref{figure2}, we compare the total strength $M_0$, which enters 
Eq.~(\ref{2}), for $\hat O=\hat D$. We evaluate $M_0$ for the dipole operator by 
means of an expansion over basis states with $J^\pi T=1^-1$. Alternatively, 
using the non-energy weighted sum rule, one can evaluate $M_0$ directly on the 
ground state as an expectation value of two long-range operators (\emph{i.e.}, 
the mean square charge radius $\langle r_{ch}^{2}\rangle$ and the mean 
square proton-proton radius $\langle r_{pp}^{2}\rangle$)~\cite{Dellafiore:1982},
\begin{equation}
{M_0=
\frac{1}{3}\left[Z^{2}\left<r_{ch}^{2}\right>-
\frac{Z(Z-1)}{2}\left<r_{pp}^{2}\right>\right]\,.}
\label{oddo}
\end{equation} 
As expected, the sum rule is satisfied for each model space.
The rate of convergence of the total strength depends upon the HO frequency, as it does 
for the g.s. energy and other observables, and the least dependence on the model 
space is obtained again for $\hbar\Omega = 19$ and $28$ MeV. Also, the best 
EIHH and NCSM results ($N_{max}$ or $K_{max}=16$) agree within $0.1\%$ for 
$\hbar\Omega=12,19$ and $28$ MeV, 
whereas one finds a larger deviation ($0.8\%$) for $\hbar\Omega=40$ 
MeV, not yet completely convergent. 
The origin of this behavior is related to the fall-off of the wave function for the HO potential 
well. As the HO potential well becomes steeper and steeper, the wave functions 
in a fixed model space go faster to zero; as a consequence, one finds smaller
root-mean square radii for the $^4$He g.s., and the expectation value of a long-range
operator, like the isovector dipole, is more poorly represented.
In principle, we could improve the convergence of $M_0$ for $\hbar\Omega=40$ MeV 
by calculating the expectation value of $D^\dagger D$ directly on the ground state up to
$N_{max}=20$. We would like to point out that, despite this, the discrepancy never exceeds the $1\%$.

\begin{figure}
\includegraphics*[scale=0.65]{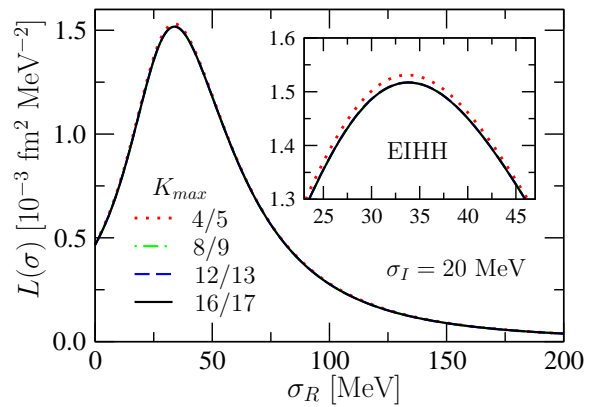}
\caption{(Color online) The LIT [see Eq.~(\ref{2})] for the isovector 
dipole transition as a function of $\sigma_R$ for $\sigma_I=20$ MeV: 
convergence of the EIHH calculation with respect to the maximal 
value of the HH grand-angular momentum quantum numbers 
$K_{max}$/$(K_{max}+1)$ in the EIHH expansions.}
\label{figure3}
\end{figure}

\begin{figure}
\includegraphics*[scale=0.65]{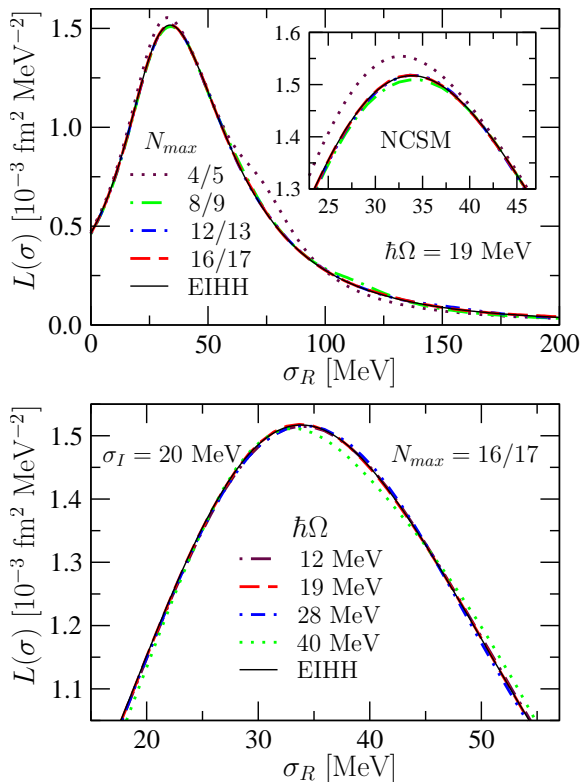}
\caption{(Color online) The LIT [see Eq.~(\ref{2})] for the isovector
dipole transition as a function of $\sigma_R$ at $\sigma_I=20$ MeV:
convergence with respect to the HO excitations allowed in the NCSM
model space for $\Omega=19$ MeV (upper panel), and dependence on the HO
frequency $\Omega$ of the best ($N_{max}=16/17$) NCSM results
(lower panel). The solid line represents the EIHH result for 
$K_{max}=16/17$.} 
\label{figure4}
\end{figure}

The overall convergence behavior of the LIT [see Eq.~(\ref{2})] 
is influenced by the convergence of its two components: (i)
the total strength $M_0$, which was discussed earlier; 
(ii) the residual continued fraction of Lanczos coefficients. 
While the first is governed by the convergence of the ground state 
(non-energy weighted sum rule), the results for the continued 
fraction have a further dependence on the $J^\pi T=1^- 1$ model space.
For this reason, the convergence pattern of the LIT has to be studied 
as a function of the model space for the ground state as well as for 
the LIT state $|\widetilde{\Psi}\rangle$. Due to the parity difference
of the two states, the values of $K_{max}$ and $N_{max}$ chosen for the 
model spaces are even for the ground state and odd for $J^\pi=1^-1$ state, 
respectively.

Figure~\ref{figure3} shows that the convergence of the LIT for 
EIHH is fast and accurate over the entire range of $\sigma_R$.
The same plot for the NCSM results at $\hbar\Omega=19$ MeV, presented
in the upper panel of Fig.~\ref{figure3}, reveals a slower convergence
rate. In particular, we notice that, unlike for the EIHH,
the high $\sigma_R$ tails of the NCSM results show oscillations and
a satisfactory quenching of such oscillations is obtained only for
the biggest model space ($N_{max}=16/17$). Furthermore, one finds a
stronger dependence on the model-space size, even in the peak region
(see the insets of Figs.~\ref{figure3} and~\ref{figure4}).

\begin{figure}
\includegraphics*[scale=0.65]{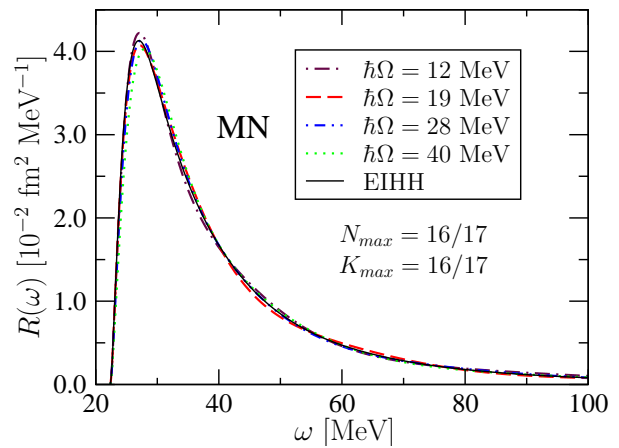}
\caption{(Color online) The NCSM ($N_{max}=16/17$) inclusive response 
to the isovector dipole transition [see Eq.~(\ref{1})] as a function 
of $\omega$ for four different values of the HO parameter $\Omega$: 
comparison with the EIHH ($K_{max}=16/17$) result.} 
\label{figure5}
\end{figure}

The influence of
the HO frequency on the LIT is shown in the lower panel
of Fig.~\ref{figure4}. In a given model space, for smaller HO 
frequencies one obtains a better
sampling of the complex-energy continuum, and therefore a more accurate convergence
of the LIT, especially in the low-$\sigma_R$ region; on the 
other hand, faster convergence is achieved when the HO length
parameter is chosen close to the size of the state to be
described, and, therefore, larger $\hbar\Omega$ values are preferrable
for describing the ground state.
As a consequence, in the particular case of the $^4$He nucleus,
frequencies in the range $12{\rm MeV}\leq\hbar\Omega\leq28{\rm MeV}$
represent a good compromise, as shown in Fig~\ref{figure4} (lower panel). 
Among the four frequencies adopted to evaluate the LIT with
the NCSM, $\hbar\Omega=19$ MeV leads to the smoothest convergence
pattern and to the best agreement with the EIHH curve, which
appears as a solid line in Fig.~\ref{figure4}. 

The different behavior of the two methods with respect to the
size of the $P$-space used in the calculation is 
related both to the differences in the shape and asymptotic
behavior of the HO and HH base states~\cite{FE:1981,barnea-1999-650}, and, more importantly,
to the actual number of states, included in the expansion.
We would like to point out that, due to the extra flexibility of the HH basis
in the hyperradial part of the expansion, for the same value of $N_{max}$ 
and $K_{max}$, the number of totally antisymmetric
basis states in the two approaches is significantly different.
For example, in the model space defined by $N_{max}=16$, one
has $2,775$ four-body states for a NCSM calculation,
while for $K_{max}=16$, the total number of four-body
states in a EIHH calculation is $10,890$. Further details on the
connection between the number of basis states in the two approaches
can be found in Refs. \cite{barnea:054001} and~\cite{barnea:p1458}.

Figure~\ref{figure5} shows the EIHH and NCSM
results for the inclusive response to the isovector dipole
excitation obtained by inverting the LIT's of Figs.~\ref{figure3}
and~\ref{figure4}. (Note that the same results for the response 
are obtained by inverting the LIT for $\sigma_I=10$ and $15$ MeV.)
The HO frequency value of 
$\hbar\Omega=19$ MeV leads to the best agreement with 
the EIHH response. Indeed, the discrepancy between the two numerical results
does not exceed $5\%$ in the energy intervals immediately close to the 
disintegration threshold as well as in the dipole resonance peak region.
While the resonant peak is equally well described for $\hbar\Omega=12$ 
and $28$ MeV, these HO frequencies yield a larger discrepancy 
(at most $10\%$ and $15\%$, respectively) 
within $3$ MeV from threshold. As expected from previous observations, poorest 
agreement is found for $\hbar\Omega=40$ MeV, although mainly in the low-energy part of
the response. In the range $60$ MeV$\leq\omega\leq80$ MeV all four NCSM responses 
 agree within the $7\%$ or better of the EIHH result.

\subsection{Quadrupole response}

\begin{figure}
\includegraphics*[scale=0.65]{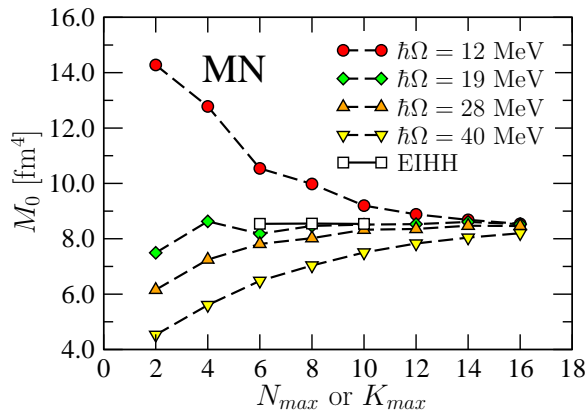}
\caption{(Color online) $^4$He total quadrupole strength for 
the MN potential as a function of the HO excitations allowed in 
the NCSM model space and of the maximal value of the HH grand-angular 
momentum quantum number $K_{max}$ in the EIHH expansion.} 
\label{figureQ}
\end{figure}

\begin{figure}
\includegraphics*[scale=0.65]{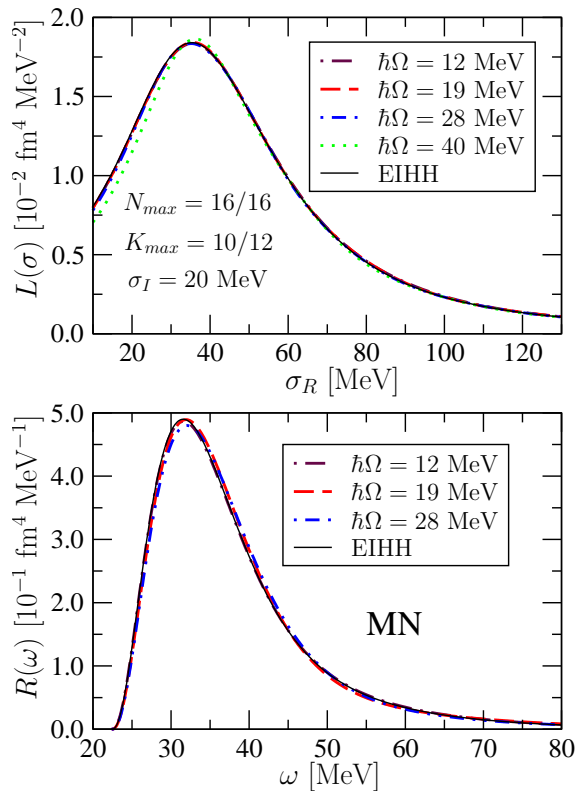}
\caption{(Color online) The dependence on the HO frequency $\Omega$ 
of (upper panel) the LIT for $\sigma_I=20$ MeV, and (lower panel) the inclusive response  
for the isoscalar quadrupole transition. Note that $N_{max}=16/16$ 
is not a complete model space (see text for details). 
The solid lines represent the EIHH results for $K_{max}=10/12$.} 
\label{figureQ2}
\end{figure}

We present now the results for the isoscalar quadrupole transition.
As expected, the total 
strength $M_0$ for the isoscalar quadrupole, shown in Fig.~\ref{figureQ},
presents a slower convergence rate than the same quantity for
the dipole operator. The greatest sensitivity to the model-space size is obtained
again for $\hbar\Omega=40$ MeV. The calculation of the total strength
for the quadrupole transition employs an expansion over an intermediate
set of basis states with $J^\pi T=2^+0$. In particular, for the NCSM
values in the largest model space ($N_{max}=16/16$), we omit the
contribution of the transitions to the $J^\pi T=2^+0$ states
with $N_{max}=18$, as a full calculation in this case would be out of reach.
In principle, one could evaluate the loss of strength induced by
such an approximation (introduced due to computational limitations),
using the non-energy weighted sum rule and calculating the expectation
value of the operator $\hat{Q}^\dagger\hat Q$ directly on the ground
state, as was done for the dipole response. However, the form of
this operator makes the calculation technically difficult.
As for the dipole transition, the EIHH calculation shows a fast
convergence. The agreement with the NCSM results is within the
1\% for $\hbar\Omega=12,19$ and $28$ MeV, while $\hbar\Omega=40$ MeV,
for which the convergence is not completely reached, shows a 5\%
discrepancy. Hence, we can assume that the transitions to $N_{max}=18$ can
be neglected.

The comparison of the LIT's in the biggest model space 
($N_{max}=16/16$, $K_{max}=10/12$) for the EIHH and the four 
different NCSM calculations is presented in the upper panel 
of Fig.~\ref{figureQ2}. All the curves, excluding the result 
for $\hbar\Omega=40$ MeV, still far from convergence, show a 
good agreement, and the corresponding inversions are compared 
in the lower panel of Fig.~\ref{figureQ2}. The discrepancies 
in the response functions do not 
exceed the 10\% in the range $25~{\rm MeV}\le\omega\le60~{\rm MeV}$, 
where the curve shows a resonant behavior. More delicate are 
the regions close to threshold and high-energy tail, 
where the response is small and the inversion procedure is 
more sensitive to the numerical error present in the LIT. 
One finds the best agreement for $\hbar\Omega=12$ MeV.

\section{Conclusions and outlook}
\label{Sec:Conclusions}
We have presented the results of an \emph{ab initio} calculation 
of the $^4$He response functions to the isovector dipole and 
isoscalar quadrupole excitations, obtained by means of the LIT 
method, within both the NCSM and the EIHH approaches. 
As NN interaction, we have used the semirealistic MN potential model. 
The aim has been to investigate the reliability of the NCSM to 
the description of inclusive response functions via the the 
integral transform method with a Lorentzian kernel.
The chosen transitions have allowed us to perform the test 
for two operators different in isospin nature and range.
In both cases our model study has shown that the NCSM can 
be successfully applied to the solution of the bound-state-like 
equations required by the LIT methods. However, due to differences
in the asymptotics of the wave functions and in the strength distribution 
in the continuum achieved with the HO and HH expansions, 
the practical implementation of the 
method, especially concerning the problems of convergence, might 
lead to difficulties. In particular, to ensure a small numerical 
uncertainty in the response function, obtained 
by numerical inversion~\cite{Efros:1999fbs,andreasi:05eps}, 
one has to achieve a very good accuracy 
in the calculation of the LIT. Consequently, 
 it is necessary to find a range of HO 
frequencies, for which both the ground and the excited states of 
the system present good convergence properties. The actual choice 
of $\hbar\Omega$ depends on both the nucleus under consideration 
and the range of the transition operator. For $^4$He we find that 
frequencies in the range $12~{\rm MeV}\leq\hbar\Omega\leq28~{\rm MeV}$
have the demanded characteristics for both the isovector dipole 
and isoscalar quadrupole excitations. 

Our main conclusion is that, in this benchmarking calculation with a semirealistic 
NN interaction, the NCSM is able to reach the level of precision required for the 
description of continuum responses via the LIT method and to obtain equivalent 
results as those found with the EIHH method. Because of the ability of NCSM 
to handle heavier mass nuclei than the EIHH, the present result opens the door
for possible LIT investigations of heavier nuclei. However, the large model spaces 
needed in the NCSM calculations in order to achieve the necessary accuracy will
require considerable thought and effort, but may be assisted by the 
use of effective field-theory two- and three-body interactions, which should be softer.

Since LIT requires a good description of
the ground-state wave function, and, at the same time, a fine enough
discretization of the complex-energy continuum, one could use a mixed-mode approach in the
spirit of Ref.~\cite{Gueorguiev:2001va}. Thus, one could use basis 
states with a frequency that ensures fast convergence for the ground
state, adding also basis states of smaller frequencies, which
would allow a better discretization of the complex-energy continuum at low energies,
and basis states corresponding to larger frequencies for a better
description of the response at higher energies. This approach
would require, however, a more involved numerical technique which can
be adapted for our no-core calculations in a Jacobi basis. Moreover,
such a method is not immediately applicable to Slater determinant
basis codes, which are much more efficient for calculations in heavier nuclei,
because of the more delicate treatment of the center-of-mass
motion required in such a case. While such an approximation is under consideration, we leave its
possible implementation for future work.
\\

 \bigskip

\begin{acknowledgments}
I.S., S.Q., and B.R.B acknowledge partial support by NFS grants PHY0070858 and PHY0244389. The work was performed in part under the auspices of the U. S. Department of Energy by the University of California, Lawrence Livermore National Laboratory under contract No. W-7405-Eng-48. P.N. received support from LDRD contract 04-ERD-058. W.L. and G.O. acknowledge support by the grant COFIN03 of the Italian Ministry of University and Research. The work of N.B. was supported by the ISRAEL SCIENCE FOUNDATION (Grant No. 361/05). C.W.J. acknowledges USDOE grant No.DE-FG02-03ER41272. We thank the Institute for Nuclear Theory at the University of Washington for its hospitality and the Department of Energy for partial support during the development of this work.
\end{acknowledgments}

\begin{thebibliography}{60}
\expandafter\ifx\csname natexlab\endcsname\relax\def\natexlab#1{#1}\fi
\expandafter\ifx\csname bibnamefont\endcsname\relax
  \def\bibnamefont#1{#1}\fi
\expandafter\ifx\csname bibfnamefont\endcsname\relax
  \def\bibfnamefont#1{#1}\fi
\expandafter\ifx\csname citenamefont\endcsname\relax
  \def\citenamefont#1{#1}\fi
\expandafter\ifx\csname url\endcsname\relax
  \def\url#1{\texttt{#1}}\fi
\expandafter\ifx\csname urlprefix\endcsname\relax\def\urlprefix{URL }\fi
\providecommand{\bibinfo}[2]{#2}
\providecommand{\eprint}[2][]{\url{#2}}

\bibitem[{\citenamefont{Arenh{\"o}vel and Sanzone}(1991)}]{AS:1991}
\bibinfo{author}{\bibfnamefont{H.}~\bibnamefont{Arenh{\"o}vel}}
  \bibnamefont{and} \bibinfo{author}{\bibfnamefont{M.}~\bibnamefont{Sanzone}},
  \bibinfo{journal}{Few. Body Syst. Suppl.} \textbf{\bibinfo{volume}{3}},
  \bibinfo{pages}{1} (\bibinfo{year}{1991}).

\bibitem[{\citenamefont{Carlson and Schiavilla}(1998)}]{Carlson:1998}
\bibinfo{author}{\bibfnamefont{J.}~\bibnamefont{Carlson}} \bibnamefont{and}
  \bibinfo{author}{\bibfnamefont{R.}~\bibnamefont{Schiavilla}},
  \bibinfo{journal}{Rev. Mod. Phys.} \textbf{\bibinfo{volume}{70}},
  \bibinfo{pages}{743} (\bibinfo{year}{1998}).

\bibitem[{\citenamefont{Arenh{\"o}vel et~al.}(2005)\citenamefont{Arenh{\"o}vel,
  Leidemann, and Tomusiak}}]{Arenhovel:2005}
\bibinfo{author}{\bibfnamefont{H.}~\bibnamefont{Arenh{\"o}vel}},
  \bibinfo{author}{\bibfnamefont{W.}~\bibnamefont{Leidemann}},
  \bibnamefont{and} \bibinfo{author}{\bibfnamefont{E.~L.}
  \bibnamefont{Tomusiak}}, \bibinfo{journal}{Eur. Phys. J}
  \textbf{\bibinfo{volume}{A23}}, \bibinfo{pages}{147} (\bibinfo{year}{2005}),
  \eprint{nucl-th/0407053}.


\bibitem[{\citenamefont{Marcucci et~al.}(2005)\citenamefont{Marcucci, Viviani,
  Schiavilla, Kievsky, and Rosati}}]{marcucci:014001}
\bibinfo{author}{\bibfnamefont{L.~E.} \bibnamefont{Marcucci}},
  \bibinfo{author}{\bibfnamefont{M.}~\bibnamefont{Viviani}},
  \bibinfo{author}{\bibfnamefont{R.}~\bibnamefont{Schiavilla}},
  \bibinfo{author}{\bibfnamefont{A.}~\bibnamefont{Kievsky}}, \bibnamefont{and}
  \bibinfo{author}{\bibfnamefont{S.}~\bibnamefont{Rosati}},
  \bibinfo{journal}{Phys. Rev. C} \textbf{\bibinfo{volume}{72}},
  \bibinfo{eid}{014001} (\bibinfo{year}{2005}).

\bibitem[{\citenamefont{Skibinski et~al.}(2005)\citenamefont{Skibinski, Golak,
  Witala, Gl{\"o}ckle, Nogga, and Kamada}}]{skibinski:044002}
\bibinfo{author}{\bibfnamefont{R.}~\bibnamefont{Skibinski}},
  \bibinfo{author}{\bibfnamefont{J.}~\bibnamefont{Golak}},
  \bibinfo{author}{\bibfnamefont{H.}~\bibnamefont{Witala}},
  \bibinfo{author}{\bibfnamefont{W.}~\bibnamefont{Gl{\"o}ckle}},
  \bibinfo{author}{\bibfnamefont{A.}~\bibnamefont{Nogga}}, \bibnamefont{and}
  \bibinfo{author}{\bibfnamefont{H.}~\bibnamefont{Kamada}},
  \bibinfo{journal}{Phys. Rev. C} \textbf{\bibinfo{volume}{72}},
  \bibinfo{eid}{044002} (\bibinfo{year}{2005}).


\bibitem[{\citenamefont{Golak et~al.}(2005)\citenamefont{Golak, Skibinski,
  Witala, Gl{\``o}ckle, Nogga, and Kamada}}]{Golak:41589}
\bibinfo{author}{\bibfnamefont{J.}~\bibnamefont{Golak}},
  \bibinfo{author}{\bibfnamefont{R.}~\bibnamefont{Skibinski}},
  \bibinfo{author}{\bibfnamefont{H.}~\bibnamefont{Witala}},
  \bibinfo{author}{\bibfnamefont{W.}~\bibnamefont{Gl{\"o}ckle}},
  \bibinfo{author}{\bibfnamefont{A.}~\bibnamefont{Nogga}}, \bibnamefont{and}
  \bibinfo{author}{\bibfnamefont{H.}~\bibnamefont{Kamada}},
  \bibinfo{journal}{Phys. Rep.} \textbf{\bibinfo{volume}{415}},
  \bibinfo{pages}{89} (\bibinfo{year}{2005}).


\bibitem[{\citenamefont{Deltuva et~al.}(2005)\citenamefont{Deltuva, Fonseca,
  and Sauer}}]{deltuva:054004}
\bibinfo{author}{\bibfnamefont{A.}~\bibnamefont{Deltuva}},
  \bibinfo{author}{\bibfnamefont{A.~C.} \bibnamefont{Fonseca}},
  \bibnamefont{and} \bibinfo{author}{\bibfnamefont{P.~U.} \bibnamefont{Sauer}},
  \bibinfo{journal}{Phys. Rev. C} \textbf{\bibinfo{volume}{72}},
  \bibinfo{eid}{054004} (\bibinfo{year}{2005}).

\bibitem[{\citenamefont{Pieper et~al.}(2002)\citenamefont{Pieper, Varga, and
  Wiringa}}]{Pieper:2002ne}
\bibinfo{author}{\bibfnamefont{S.~C.} \bibnamefont{Pieper}},
  \bibinfo{author}{\bibfnamefont{K.}~\bibnamefont{Varga}}, \bibnamefont{and}
  \bibinfo{author}{\bibfnamefont{R.~B.} \bibnamefont{Wiringa}},
  \bibinfo{journal}{Phys. Rev.} \textbf{\bibinfo{volume}{C66}},
  \bibinfo{pages}{044310} (\bibinfo{year}{2002}), \eprint{nucl-th/0206061}.

\bibitem[{\citenamefont{Pieper et~al.}(2004)\citenamefont{Pieper, Wiringa, and
  Carlson}}]{pieper:054325}
\bibinfo{author}{\bibfnamefont{S.~C.} \bibnamefont{Pieper}},
  \bibinfo{author}{\bibfnamefont{R.~B.} \bibnamefont{Wiringa}},
  \bibnamefont{and} \bibinfo{author}{\bibfnamefont{J.}~\bibnamefont{Carlson}},
  \bibinfo{journal}{Phys. Rev. C} \textbf{\bibinfo{volume}{70}},
  \bibinfo{eid}{054325} (\bibinfo{year}{2004}).

\bibitem[{\citenamefont{Navr{\'a}til
  et~al.}(2000{\natexlab{a}})\citenamefont{Navr{\'a}til, Vary, and
  Barrett}}]{Navratil:2000ww}
\bibinfo{author}{\bibfnamefont{P.}~\bibnamefont{Navr{\'a}til}},
  \bibinfo{author}{\bibfnamefont{J.~P.} \bibnamefont{Vary}}, \bibnamefont{and}
  \bibinfo{author}{\bibfnamefont{B.~R.} \bibnamefont{Barrett}},
  \bibinfo{journal}{Phys. Rev. Lett.} \textbf{\bibinfo{volume}{84}},
  \bibinfo{pages}{5728} (\bibinfo{year}{2000}{\natexlab{a}}),
  \eprint{nucl-th/0004058}.

\bibitem[{\citenamefont{Vary et~al.}(2005)\citenamefont{Vary, Atramentov,
  Barrett, Hasan, Hayes, Lloyd, Mazur, Navr{\'a}til, Negoita, Nogga
  et~al.}}]{A48}
\bibinfo{author}{\bibfnamefont{J.~P.} \bibnamefont{Vary}},
  \bibinfo{author}{\bibfnamefont{O.~V.} \bibnamefont{Atramentov}},
  \bibinfo{author}{\bibfnamefont{B.~R.} \bibnamefont{Barrett}},
  \bibinfo{author}{\bibfnamefont{M.}~\bibnamefont{Hasan}},
  \bibinfo{author}{\bibfnamefont{A.~C.} \bibnamefont{Hayes}},
  \bibinfo{author}{\bibfnamefont{R.}~\bibnamefont{Lloyd}},
  \bibinfo{author}{\bibfnamefont{A.~I.} \bibnamefont{Mazur}},
  \bibinfo{author}{\bibfnamefont{P.}~\bibnamefont{Navr{\'a}til}},
  \bibinfo{author}{\bibfnamefont{A.~G.} \bibnamefont{Negoita}},
  \bibinfo{author}{\bibfnamefont{A.}~\bibnamefont{Nogga}},
  \bibnamefont{et~al.}, \bibinfo{journal}{Eur. Phys. J.}
  \textbf{\bibinfo{volume}{A25}}, \bibinfo{pages}{475} (\bibinfo{year}{2005}).

\bibitem[{ne2()}]{ne20NCSM}
\bibinfo{note}{See, e.g., $^{20}$Ne example in P. Navr{\'a}til's talk},
  \urlprefix\url{http://www.int.washington.edu/talks/WorkShops/int_05_3/}.

\bibitem[{\citenamefont{Efros}(1985)}]{EFROS:1985}
\bibinfo{author}{\bibfnamefont{V.~D.} \bibnamefont{Efros}},
  \bibinfo{journal}{Yad.\ Fiz.} \textbf{\bibinfo{volume}{41}},
  \bibinfo{pages}{1498} (\bibinfo{year}{1985}) \bibinfo{note}{[Sov. J. Nucl.
  Phys. {\bf 41}, 949 (1985)]}.

\bibitem[{\citenamefont{Efros}(1993)}]{EFROS:1993}
\bibinfo{author}{\bibfnamefont{V.~D.} \bibnamefont{Efros}},
  \bibinfo{journal}{Yad.\ Fiz.} \textbf{\bibinfo{volume}{56, N7}},
  \bibinfo{pages}{22} (\bibinfo{year}{1993}) \bibinfo{note}{[Phys. At. Nucl.
  {\bf 56}, 869 (1993)]}.

\bibitem[{\citenamefont{Efros}(1999)}]{EFROS:1999}
\bibinfo{author}{\bibfnamefont{V.~D.} \bibnamefont{Efros}},
  \bibinfo{journal}{Yad.\ Fiz.} \textbf{\bibinfo{volume}{62}},
  \bibinfo{pages}{1975} (\bibinfo{year}{1999}) \bibinfo{note}{[Phys. At. Nucl.
  {\bf 62}, 1833 (1999)]}.

\bibitem[{\citenamefont{Carlson and Schiavilla}(1992)}]{Carlson:1992}
\bibinfo{author}{\bibfnamefont{J.}~\bibnamefont{Carlson}} \bibnamefont{and}
  \bibinfo{author}{\bibfnamefont{R.}~\bibnamefont{Schiavilla}},
  \bibinfo{journal}{Phys. Rev. Lett.} \textbf{\bibinfo{volume}{68}},
  \bibinfo{pages}{3682} (\bibinfo{year}{1992}).

\bibitem[{\citenamefont{Efros et~al.}(1993)\citenamefont{Efros, Leidemann, and
  Orlandini}}]{Efros:1993fbs}
\bibinfo{author}{\bibfnamefont{V.~D.} \bibnamefont{Efros}},
  \bibinfo{author}{\bibfnamefont{W.}~\bibnamefont{Leidemann}},
  \bibnamefont{and}
  \bibinfo{author}{\bibfnamefont{G.}~\bibnamefont{Orlandini}},
  \bibinfo{journal}{Few-Body Syst.} \textbf{\bibinfo{volume}{14}},
  \bibinfo{pages}{151} (\bibinfo{year}{1993}).

\bibitem[{\citenamefont{Tikonov and Arsenin}(1977)}]{Tikonov:1977}
\bibinfo{author}{\bibfnamefont{A.~N.} \bibnamefont{Tikonov}} \bibnamefont{and}
  \bibinfo{author}{\bibfnamefont{V.~Y.} \bibnamefont{Arsenin}},
  \emph{\bibinfo{title}{Solution of Ill Posed Problems}}
  (\bibinfo{publisher}{V. H. Winston {\&} sons, Whashington D.C.},
  \bibinfo{year}{1977}).

\bibitem[{\citenamefont{Efros et~al.}(1994)\citenamefont{Efros, Leidemann, and
  Orlandini}}]{Efros:1994iq}
\bibinfo{author}{\bibfnamefont{V.~D.} \bibnamefont{Efros}},
  \bibinfo{author}{\bibfnamefont{W.}~\bibnamefont{Leidemann}},
  \bibnamefont{and}
  \bibinfo{author}{\bibfnamefont{G.}~\bibnamefont{Orlandini}},
  \bibinfo{journal}{Phys. Lett.} \textbf{\bibinfo{volume}{B338}},
  \bibinfo{pages}{130} (\bibinfo{year}{1994}), \eprint{nucl-th/9409004}.

\bibitem[{\citenamefont{Efros et~al.}(2005)\citenamefont{Efros, Leidemann,
  Orlandini, and Tomusiak}}]{Efros:011002}
\bibinfo{author}{\bibfnamefont{V.~D.} \bibnamefont{Efros}},
  \bibinfo{author}{\bibfnamefont{W.}~\bibnamefont{Leidemann}},
  \bibinfo{author}{\bibfnamefont{G.}~\bibnamefont{Orlandini}},
  \bibnamefont{and} \bibinfo{author}{\bibfnamefont{E.~L.}
  \bibnamefont{Tomusiak}}, \bibinfo{journal}{Phys. Rev. C}
  \textbf{\bibinfo{volume}{72}}, \bibinfo{pages}{011002(R)}
  (\bibinfo{year}{2005}).

\bibitem[{\citenamefont{Bacca et~al.}(2004{\natexlab{a}})\citenamefont{Bacca,
  Arenh{\"o}vel, Barnea, Leidemann, and Orlandini}}]{Bacca:2004dr}
\bibinfo{author}{\bibfnamefont{S.}~\bibnamefont{Bacca}},
  \bibinfo{author}{\bibfnamefont{H.}~\bibnamefont{Arenh{\"o}vel}},
  \bibinfo{author}{\bibfnamefont{N.}~\bibnamefont{Barnea}},
  \bibinfo{author}{\bibfnamefont{W.}~\bibnamefont{Leidemann}},
  \bibnamefont{and}
  \bibinfo{author}{\bibfnamefont{G.}~\bibnamefont{Orlandini}},
  \bibinfo{journal}{Phys. Lett.} \textbf{\bibinfo{volume}{B603}},
  \bibinfo{pages}{159} (\bibinfo{year}{2004}{\natexlab{a}}),
  \eprint{nucl-th/0406080}.

\bibitem[{\citenamefont{Barnea et~al.}(2000{\natexlab{a}})\citenamefont{Barnea,
  Leidemann, and Orlandini}}]{barnea:054001}
\bibinfo{author}{\bibfnamefont{N.}~\bibnamefont{Barnea}},
  \bibinfo{author}{\bibfnamefont{W.}~\bibnamefont{Leidemann}},
  \bibnamefont{and}
  \bibinfo{author}{\bibfnamefont{G.}~\bibnamefont{Orlandini}},
  \bibinfo{journal}{Phys. Rev. C} \textbf{\bibinfo{volume}{61}},
  \bibinfo{pages}{054001} (\bibinfo{year}{2000}{\natexlab{a}}).

\bibitem[{\citenamefont{Barnea et~al.}(2003)\citenamefont{Barnea, Leidemann,
  and Orlandini}}]{barnea:054003}
\bibinfo{author}{\bibfnamefont{N.}~\bibnamefont{Barnea}},
  \bibinfo{author}{\bibfnamefont{W.}~\bibnamefont{Leidemann}},
  \bibnamefont{and}
  \bibinfo{author}{\bibfnamefont{G.}~\bibnamefont{Orlandini}},
  \bibinfo{journal}{Phys. Rev. C} \textbf{\bibinfo{volume}{67}},
  \bibinfo{eid}{054003} (\bibinfo{year}{2003}).

\bibitem[{\citenamefont{Bacca et~al.}(2002)\citenamefont{Bacca, Marchisio,
  Barnea, Leidemann, and Orlandini}}]{Bacca:2001kr}
\bibinfo{author}{\bibfnamefont{S.}~\bibnamefont{Bacca}},
  \bibinfo{author}{\bibfnamefont{M.~A.} \bibnamefont{Marchisio}},
  \bibinfo{author}{\bibfnamefont{N.}~\bibnamefont{Barnea}},
  \bibinfo{author}{\bibfnamefont{W.}~\bibnamefont{Leidemann}},
  \bibnamefont{and}
  \bibinfo{author}{\bibfnamefont{G.}~\bibnamefont{Orlandini}},
  \bibinfo{journal}{Phys. Rev. Lett.} \textbf{\bibinfo{volume}{89}},
  \bibinfo{pages}{052502} (\bibinfo{year}{2002}), \eprint{nucl-th/0112067}.

\bibitem[{\citenamefont{Bacca et~al.}(2004{\natexlab{b}})\citenamefont{Bacca,
  Barnea, Leidemann, and Orlandini}}]{bacca:057001}
\bibinfo{author}{\bibfnamefont{S.}~\bibnamefont{Bacca}},
  \bibinfo{author}{\bibfnamefont{N.}~\bibnamefont{Barnea}},
  \bibinfo{author}{\bibfnamefont{W.}~\bibnamefont{Leidemann}},
  \bibnamefont{and}
  \bibinfo{author}{\bibfnamefont{G.}~\bibnamefont{Orlandini}},
  \bibinfo{journal}{Phys. Rev. C.} \textbf{\bibinfo{volume}{69}},
  \bibinfo{eid}{057001} (\bibinfo{year}{2004}{\natexlab{b}}).

\bibitem[{\citenamefont{Gazit et~al.}(2006)\citenamefont{Gazit, Bacca, Barnea,
  Leidemann, and Orlandini}}]{gazit:112301}
\bibinfo{author}{\bibfnamefont{D.}~\bibnamefont{Gazit}},
  \bibinfo{author}{\bibfnamefont{S.}~\bibnamefont{Bacca}},
  \bibinfo{author}{\bibfnamefont{N.}~\bibnamefont{Barnea}},
  \bibinfo{author}{\bibfnamefont{W.}~\bibnamefont{Leidemann}},
  \bibnamefont{and}
  \bibinfo{author}{\bibfnamefont{G.}~\bibnamefont{Orlandini}},
  \bibinfo{journal}{Phys. Rev. Lett.} \textbf{\bibinfo{volume}{96}},
  \bibinfo{eid}{112301} (\bibinfo{year}{2006}).

\bibitem[{\citenamefont{Quaglioni et~al.}(2004)\citenamefont{Quaglioni,
  Leidemann, Orlandini, Barnea, and Efros}}]{quaglioni:044002}
\bibinfo{author}{\bibfnamefont{S.}~\bibnamefont{Quaglioni}},
  \bibinfo{author}{\bibfnamefont{W.}~\bibnamefont{Leidemann}},
  \bibinfo{author}{\bibfnamefont{G.}~\bibnamefont{Orlandini}},
  \bibinfo{author}{\bibfnamefont{N.}~\bibnamefont{Barnea}}, \bibnamefont{and}
  \bibinfo{author}{\bibfnamefont{V.~D.} \bibnamefont{Efros}},
  \bibinfo{journal}{Phys. Rev. C} \textbf{\bibinfo{volume}{69}},
  \bibinfo{eid}{044002} (\bibinfo{year}{2004}).

\bibitem[{\citenamefont{Quaglioni et~al.}(2005)\citenamefont{Quaglioni, Efros,
  Leidemann, and Orlandini}}]{quaglioni:064002}
\bibinfo{author}{\bibfnamefont{S.}~\bibnamefont{Quaglioni}},
  \bibinfo{author}{\bibfnamefont{V.~D.} \bibnamefont{Efros}},
  \bibinfo{author}{\bibfnamefont{W.}~\bibnamefont{Leidemann}},
  \bibnamefont{and}
  \bibinfo{author}{\bibfnamefont{G.}~\bibnamefont{Orlandini}},
  \bibinfo{journal}{Phys. Rev. C} \textbf{\bibinfo{volume}{72}},
  \bibinfo{eid}{064002} (\bibinfo{year}{2005}).

\bibitem[{\citenamefont{Andreasi et~al.}(2006)\citenamefont{Andreasi,
  Quaglioni, Efros, Leidemann, and Orlandini}}]{andreasi:06eps}
\bibinfo{author}{\bibfnamefont{D.}~\bibnamefont{Andreasi}},
  \bibinfo{author}{\bibfnamefont{S.}~\bibnamefont{Quaglioni}},
  \bibinfo{author}{\bibfnamefont{V.~D.} \bibnamefont{Efros}},
  \bibinfo{author}{\bibfnamefont{W.}~\bibnamefont{Leidemann}},
  \bibnamefont{and}
  \bibinfo{author}{\bibfnamefont{G.}~\bibnamefont{Orlandini}},
  \bibinfo{journal}{Eur. Phys. J} \textbf{\bibinfo{volume}{A27}},
  \bibinfo{pages}{47} (\bibinfo{year}{2006}).

\bibitem[{\citenamefont{Thomson et~al.}(1977)\citenamefont{Thomson, LeMere, and
  Tang}}]{Minnesota}
\bibinfo{author}{\bibfnamefont{D.~R.} \bibnamefont{Thomson}},
  \bibinfo{author}{\bibfnamefont{M.}~\bibnamefont{LeMere}}, \bibnamefont{and}
  \bibinfo{author}{\bibfnamefont{Y.~C.} \bibnamefont{Tang}},
  \bibinfo{journal}{Nucl. Phys.} \textbf{\bibinfo{volume}{A286}},
  \bibinfo{pages}{53} (\bibinfo{year}{1977}).

\bibitem[{\citenamefont{Marchisio et~al.}(2003)\citenamefont{Marchisio, Barnea,
  Leidemann, and Orlandini}}]{Marchisio:2003}
\bibinfo{author}{\bibfnamefont{M.~A.} \bibnamefont{Marchisio}},
  \bibinfo{author}{\bibfnamefont{N.}~\bibnamefont{Barnea}},
  \bibinfo{author}{\bibfnamefont{W.}~\bibnamefont{Leidemann}},
  \bibnamefont{and}
  \bibinfo{author}{\bibfnamefont{G.}~\bibnamefont{Orlandini}},
  \bibinfo{journal}{Few-Body Syst.} \textbf{\bibinfo{volume}{33}},
  \bibinfo{pages}{259} (\bibinfo{year}{2003}), \eprint{nucl-th/0202009}.

\bibitem[{\citenamefont{Efros et~al.}(1999)\citenamefont{Efros, Leidemann, and
  Orlandini}}]{Efros:1999fbs}
\bibinfo{author}{\bibfnamefont{V.~D.} \bibnamefont{Efros}},
  \bibinfo{author}{\bibfnamefont{W.}~\bibnamefont{Leidemann}},
  \bibnamefont{and}
  \bibinfo{author}{\bibfnamefont{G.}~\bibnamefont{Orlandini}},
  \bibinfo{journal}{Few-Body Syst.} \textbf{\bibinfo{volume}{26}},
  \bibinfo{pages}{251} (\bibinfo{year}{1999}).

\bibitem[{\citenamefont{Andreasi et~al.}(2005)\citenamefont{Andreasi,
  Leidemann, Rei{\ss}, and Schwamb}}]{andreasi:05eps}
\bibinfo{author}{\bibfnamefont{D.}~\bibnamefont{Andreasi}},
  \bibinfo{author}{\bibfnamefont{W.}~\bibnamefont{Leidemann}},
  \bibinfo{author}{\bibfnamefont{C.}~\bibnamefont{Rei{\ss}}}, \bibnamefont{and}
  \bibinfo{author}{\bibfnamefont{M.}~\bibnamefont{Schwamb}},
  \bibinfo{journal}{Eur. Phys. J} \textbf{\bibinfo{volume}{A24}},
  \bibinfo{pages}{361} (\bibinfo{year}{2005}).

\bibitem[{\citenamefont{Caurier et~al.}(1990)\citenamefont{Caurier, Zuker, and
  Poves}}]{Caurier:1990dc}
\bibinfo{author}{\bibfnamefont{E.}~\bibnamefont{Caurier}},
  \bibinfo{author}{\bibfnamefont{A.~P.} \bibnamefont{Zuker}}, \bibnamefont{and}
  \bibinfo{author}{\bibfnamefont{A.}~\bibnamefont{Poves}},
  \bibinfo{journal}{Phys. Lett.} \textbf{\bibinfo{volume}{B252}},
  \bibinfo{pages}{13} (\bibinfo{year}{1990}).

\bibitem[{\citenamefont{Caurier et~al.}(1995)\citenamefont{Caurier, Poves, and
  Zuker}}]{Caurier:1994xg}
\bibinfo{author}{\bibfnamefont{E.}~\bibnamefont{Caurier}},
  \bibinfo{author}{\bibfnamefont{A.}~\bibnamefont{Poves}}, \bibnamefont{and}
  \bibinfo{author}{\bibfnamefont{A.~P.} \bibnamefont{Zuker}},
  \bibinfo{journal}{Phys. Rev. Lett.} \textbf{\bibinfo{volume}{74}},
  \bibinfo{pages}{1517} (\bibinfo{year}{1995}), \eprint{nucl-th/9401010}.

\bibitem[{\citenamefont{Haxton et~al.}(2005)\citenamefont{Haxton, Nollett, and
  Zurek}}]{haxton:065501}
\bibinfo{author}{\bibfnamefont{W.~C.} \bibnamefont{Haxton}},
  \bibinfo{author}{\bibfnamefont{K.~M.} \bibnamefont{Nollett}},
  \bibnamefont{and} \bibinfo{author}{\bibfnamefont{K.~M.} \bibnamefont{Zurek}},
  \bibinfo{journal}{Phys. Rev. C} \textbf{\bibinfo{volume}{72}},
  \bibinfo{eid}{065501} (\bibinfo{year}{2005}).

\bibitem[{\citenamefont{Navr{\'a}til and Barrett}(1996)}]{Navratil:1996vm}
\bibinfo{author}{\bibfnamefont{P.}~\bibnamefont{Navr{\'a}til}}
  \bibnamefont{and} \bibinfo{author}{\bibfnamefont{B.~R.}
  \bibnamefont{Barrett}}, \bibinfo{journal}{Phys. Rev. C}
  \textbf{\bibinfo{volume}{54}}, \bibinfo{pages}{2986} (\bibinfo{year}{1996}),
  \eprint{nucl-th/9609046}.

\bibitem[{\citenamefont{Navr{\'a}til and Barrett}(1998)}]{Navratil:1997qb}
\bibinfo{author}{\bibfnamefont{P.}~\bibnamefont{Navr{\'a}til}}
  \bibnamefont{and} \bibinfo{author}{\bibfnamefont{B.~R.}
  \bibnamefont{Barrett}}, \bibinfo{journal}{Phys. Rev. C}
  \textbf{\bibinfo{volume}{57}}, \bibinfo{pages}{562} (\bibinfo{year}{1998}),
  \eprint{nucl-th/9711027}.

\bibitem[{\citenamefont{Da~Providencia and Shakin}(1964)}]{DaProvidencia:1964}
\bibinfo{author}{\bibfnamefont{J.}~\bibnamefont{Da~Providencia}}
  \bibnamefont{and} \bibinfo{author}{\bibfnamefont{C.~M.}
  \bibnamefont{Shakin}}, \bibinfo{journal}{Ann. of Phys.}
  \textbf{\bibinfo{volume}{30}}, \bibinfo{pages}{95} (\bibinfo{year}{1964}).

\bibitem[{\citenamefont{Suzuki and Lee}(1980)}]{Suzuki:1980}
\bibinfo{author}{\bibfnamefont{K.}~\bibnamefont{Suzuki}} \bibnamefont{and}
  \bibinfo{author}{\bibfnamefont{S.}~\bibnamefont{Lee}},
  \bibinfo{journal}{Prog. Theor. Phys.} \textbf{\bibinfo{volume}{64}},
  \bibinfo{pages}{2091} (\bibinfo{year}{1980}).

\bibitem[{\citenamefont{Suzuki}(1982)}]{Suzuki:1982}
\bibinfo{author}{\bibfnamefont{K.}~\bibnamefont{Suzuki}},
  \bibinfo{journal}{Prog. Theor. Phys.} \textbf{\bibinfo{volume}{68}},
  \bibinfo{pages}{246} (\bibinfo{year}{1982}).

\bibitem[{\citenamefont{Suzuki and Okamoto}(1983)}]{Suzuki:1983}
\bibinfo{author}{\bibfnamefont{K.}~\bibnamefont{Suzuki}} \bibnamefont{and}
  \bibinfo{author}{\bibfnamefont{R.}~\bibnamefont{Okamoto}},
  \bibinfo{journal}{Prog. Theor. Phys.} \textbf{\bibinfo{volume}{70}},
  \bibinfo{pages}{439} (\bibinfo{year}{1983}).

\bibitem[{\citenamefont{Mintkevich and Barnea}(2004)}]{mintkevich:044005}
\bibinfo{author}{\bibfnamefont{O.}~\bibnamefont{Mintkevich}} \bibnamefont{and}
  \bibinfo{author}{\bibfnamefont{N.}~\bibnamefont{Barnea}},
  \bibinfo{journal}{Phys. Rev. C} \textbf{\bibinfo{volume}{69}},
  \bibinfo{eid}{044005} (\bibinfo{year}{2004}).

\bibitem[{\citenamefont{Navr{\'a}til
  et~al.}(2000{\natexlab{b}})\citenamefont{Navr{\'a}til, Kamuntavicius, and
  Barrett}}]{Navratil:1999pw}
\bibinfo{author}{\bibfnamefont{P.}~\bibnamefont{Navr{\'a}til}},
  \bibinfo{author}{\bibfnamefont{G.~P.} \bibnamefont{Kamuntavicius}},
  \bibnamefont{and} \bibinfo{author}{\bibfnamefont{B.~R.}
  \bibnamefont{Barrett}}, \bibinfo{journal}{Phys. Rev. C}
  \textbf{\bibinfo{volume}{61}}, \bibinfo{pages}{044001}
  (\bibinfo{year}{2000}{\natexlab{b}}), \eprint{nucl-th/9907054}.

\bibitem[{\citenamefont{Navr{\'a}til
  et~al.}(2000{\natexlab{c}})\citenamefont{Navr{\'a}til, Vary, and
  Barrett}}]{Navratil:2000gs}
\bibinfo{author}{\bibfnamefont{P.}~\bibnamefont{Navr{\'a}til}},
  \bibinfo{author}{\bibfnamefont{J.~P.} \bibnamefont{Vary}}, \bibnamefont{and}
  \bibinfo{author}{\bibfnamefont{B.~R.} \bibnamefont{Barrett}},
  \bibinfo{journal}{Phys. Rev. C} \textbf{\bibinfo{volume}{62}},
  \bibinfo{pages}{054311} (\bibinfo{year}{2000}{\natexlab{c}}).

\bibitem[{\citenamefont{Nogga et~al.}(2005)\citenamefont{Nogga, Navr{\'a}til,
  Barrett, and Vary}}]{Nogga:2005hp}
\bibinfo{author}{\bibfnamefont{A.}~\bibnamefont{Nogga}},
  \bibinfo{author}{\bibfnamefont{P.}~\bibnamefont{Navr{\'a}til}},
  \bibinfo{author}{\bibfnamefont{B.~R.} \bibnamefont{Barrett}},
  \bibnamefont{and} \bibinfo{author}{\bibfnamefont{J.~P.} \bibnamefont{Vary}}
  (\bibinfo{year}{2005}), \eprint{nucl-th/0511082}.

\bibitem[{\citenamefont{Navr{\'a}til and Ormand}(2002)}]{Navratil:152502}
\bibinfo{author}{\bibfnamefont{P.}~\bibnamefont{Navr{\'a}til}}
  \bibnamefont{and} \bibinfo{author}{\bibfnamefont{W.~E.}
  \bibnamefont{Ormand}}, \bibinfo{journal}{Phys. Rev. Lett.}
  \textbf{\bibinfo{volume}{88}}, \bibinfo{pages}{152502}
  (\bibinfo{year}{2002}); \bibinfo{note}{{Phys. Rev. C {\bf 68}, 034305
  (2003)}}.

\bibitem[{\citenamefont{Stetcu et~al.}(2005)\citenamefont{Stetcu, Barrett,
  Navr\'atil, and Vary}}]{Stetcu:2004wh}
\bibinfo{author}{\bibfnamefont{I.}~\bibnamefont{Stetcu}},
  \bibinfo{author}{\bibfnamefont{B.~R.} \bibnamefont{Barrett}},
  \bibinfo{author}{\bibfnamefont{P.}~\bibnamefont{Navr\'atil}},
  \bibnamefont{and} \bibinfo{author}{\bibfnamefont{J.~P.} \bibnamefont{Vary}},
  \bibinfo{journal}{Phys. Rev. C} \textbf{\bibinfo{volume}{71}},
  \bibinfo{pages}{044325} (\bibinfo{year}{2005}), \eprint{nucl-th/0412004}.

\bibitem[{\citenamefont{Stetcu et~al.}(2006)\citenamefont{Stetcu, Barrett,
  Navratil, and Vary}}]{stetcu:037307}
\bibinfo{author}{\bibfnamefont{I.}~\bibnamefont{Stetcu}},
  \bibinfo{author}{\bibfnamefont{B.~R.} \bibnamefont{Barrett}},
  \bibinfo{author}{\bibfnamefont{P.}~\bibnamefont{Navratil}}, \bibnamefont{and}
  \bibinfo{author}{\bibfnamefont{J.~P.} \bibnamefont{Vary}},
  \bibinfo{journal}{Phys. Rev. C} \textbf{\bibinfo{volume}{73}},
  \bibinfo{eid}{037307} (\bibinfo{year}{2006}).

\bibitem[{\citenamefont{Efros}(1972)}]{Efros:1972}
\bibinfo{author}{\bibfnamefont{V.~D.} \bibnamefont{Efros}},
  \bibinfo{journal}{Yad.\ Fiz.} \textbf{\bibinfo{volume}{15}},
  \bibinfo{pages}{226} (\bibinfo{year}{1972}) \bibinfo{note}{{[}Sov. J. Nucl.
  Phys. {\bf 15}, 128 (1972){]}}.

\bibitem[{\citenamefont{Barnea and Novoselsky}(1997)}]{Barnea:1997}
\bibinfo{author}{\bibfnamefont{N.}~\bibnamefont{Barnea}} \bibnamefont{and}
  \bibinfo{author}{\bibfnamefont{A.}~\bibnamefont{Novoselsky}},
  \bibinfo{journal}{Ann. Phys. (N.Y.)} \textbf{\bibinfo{volume}{256}},
  \bibinfo{pages}{192} (\bibinfo{year}{1997}).

\bibitem[{\citenamefont{Barnea and Novoselsky}(1998)}]{Barnea:1998}
\bibinfo{author}{\bibfnamefont{N.}~\bibnamefont{Barnea}} \bibnamefont{and}
  \bibinfo{author}{\bibfnamefont{A.}~\bibnamefont{Novoselsky}},
  \bibinfo{journal}{Phys. Rev. A} \textbf{\bibinfo{volume}{57}},
  \bibinfo{pages}{48} (\bibinfo{year}{1998}).

\bibitem[{\citenamefont{Novoselsky and Katriel}(1994)}]{Novoselsky:1994}
\bibinfo{author}{\bibfnamefont{A.}~\bibnamefont{Novoselsky}} \bibnamefont{and}
  \bibinfo{author}{\bibfnamefont{J.}~\bibnamefont{Katriel}},
  \bibinfo{journal}{Phys. Rev. A} \textbf{\bibinfo{volume}{49}},
  \bibinfo{pages}{833} (\bibinfo{year}{1994}).

\bibitem[{\citenamefont{Novoselsky and Barnea}(1995)}]{Novoselsky:1995}
\bibinfo{author}{\bibfnamefont{A.}~\bibnamefont{Novoselsky}} \bibnamefont{and}
  \bibinfo{author}{\bibfnamefont{N.}~\bibnamefont{Barnea}},
  \bibinfo{journal}{Phys. Rev. A} \textbf{\bibinfo{volume}{51}},
  \bibinfo{pages}{2777} (\bibinfo{year}{1995}).

\bibitem[{\citenamefont{Barnea}(1999)}]{Barnea:1999}
\bibinfo{author}{\bibfnamefont{N.}~\bibnamefont{Barnea}},
  \bibinfo{journal}{Phys. Rev. A} \textbf{\bibinfo{volume}{59}},
  \bibinfo{pages}{1135} (\bibinfo{year}{1999}).

\bibitem[{\citenamefont{Barnea et~al.}(2000{\natexlab{b}})\citenamefont{Barnea,
  Leidemann, and Orlandini}}]{Barnea:1999be}
\bibinfo{author}{\bibfnamefont{N.}~\bibnamefont{Barnea}},
  \bibinfo{author}{\bibfnamefont{W.}~\bibnamefont{Leidemann}},
  \bibnamefont{and}
  \bibinfo{author}{\bibfnamefont{G.}~\bibnamefont{Orlandini}},
  \bibinfo{journal}{Phys. Rev. C} \textbf{\bibinfo{volume}{61}},
  \bibinfo{pages}{054001} (\bibinfo{year}{2000}{\natexlab{b}}),
  \eprint{nucl-th/9910062}.

\bibitem[{\citenamefont{Dellafiore and Lipparini}(1982)}]{Dellafiore:1982}
\bibinfo{author}{\bibfnamefont{A.}~\bibnamefont{Dellafiore}} \bibnamefont{and}
  \bibinfo{author}{\bibfnamefont{E.}~\bibnamefont{Lipparini}},
  \bibinfo{journal}{Nucl. Phys.} \textbf{\bibinfo{volume}{A388}},
  \bibinfo{pages}{639} (\bibinfo{year}{1982}).

\bibitem[{\citenamefont{Fomin and Efros}(1981)}]{FE:1981}
\bibinfo{author}{\bibfnamefont{B.~A.} \bibnamefont{Fomin}} \bibnamefont{and}
  \bibinfo{author}{\bibfnamefont{V.~D.} \bibnamefont{Efros}},
  \bibinfo{journal}{Yad.\ Fiz.} \textbf{\bibinfo{volume}{34}},
  \bibinfo{pages}{587} (\bibinfo{year}{1981}) \bibinfo{note}{[Sov. J. Nucl.
  Phys. {\bf 34}, 327 (1981)]}.

\bibitem[{\citenamefont{Barnea et~al.}(1999)\citenamefont{Barnea, Leidemann,
  and Orlandini}}]{barnea-1999-650}
\bibinfo{author}{\bibfnamefont{N.}~\bibnamefont{Barnea}},
  \bibinfo{author}{\bibfnamefont{W.}~\bibnamefont{Leidemann}},
  \bibnamefont{and}
  \bibinfo{author}{\bibfnamefont{G.}~\bibnamefont{Orlandini}},
  \bibinfo{journal}{Nucl. Phys.} \textbf{\bibinfo{volume}{A650}},
  \bibinfo{pages}{427} (\bibinfo{year}{1999}).

\bibitem[{\citenamefont{Barnea and Mandelzweig}(1992)}]{barnea:p1458}
\bibinfo{author}{\bibfnamefont{N.}~\bibnamefont{Barnea}} \bibnamefont{and}
  \bibinfo{author}{\bibfnamefont{V.~B.} \bibnamefont{Mandelzweig}},
  \bibinfo{journal}{Phys. Rev. C} \textbf{\bibinfo{volume}{45}},
  \bibinfo{pages}{1458} (\bibinfo{year}{1992}).

\bibitem[{\citenamefont{Gueorguiev et~al.}(2002)\citenamefont{Gueorguiev,
  Ormand, Johnson, and Draayer}}]{Gueorguiev:2001va}
\bibinfo{author}{\bibfnamefont{V.~G.} \bibnamefont{Gueorguiev}},
  \bibinfo{author}{\bibfnamefont{W.~E.} \bibnamefont{Ormand}},
  \bibinfo{author}{\bibfnamefont{C.~W.} \bibnamefont{Johnson}},
  \bibnamefont{and} \bibinfo{author}{\bibfnamefont{J.~P.}
  \bibnamefont{Draayer}}, \bibinfo{journal}{Phys. Rev. C}
  \textbf{\bibinfo{volume}{65}}, \bibinfo{pages}{024314}
  (\bibinfo{year}{2002}), \eprint{nucl-th/0110047}.

\end{thebibliography}

\end{document}